\documentclass[pra,aps,superscriptaddress]{revtex4}
\usepackage{graphicx}
\usepackage{amsmath}
\usepackage{epsfig}
\usepackage{amssymb}
\usepackage{epstopdf}
\begin{document}
\title{Universal entangler with photon pairs in arbitrary states}

\author{Bing He}
\email{bhe98@earthlink.net}
\address{Institute for Quantum Information Science, University of Calgary, Alberta T2N
1N4, Canada}
\address{Department of Physics and Astronomy, Hunter College of the City University
of New York, 695 Park Avenue, New York, NY 10065}
\author{Yu-Hang Ren}
\address{Department of Physics and Astronomy, Hunter College of the City University
of New York, 695 Park Avenue, New York, NY 10065}
\author{J\'{a}nos A. Bergou}
\address{Department of Physics and Astronomy, Hunter College of the City University
of New York, 695 Park Avenue, New York, NY 10065}

\begin{abstract}
We propose a setup that transforms a photon pair in arbitrary rank-four mixed state, which could also be unknown, to a Bell state. The setup involves two linear optical circuits processing the individual photons and a parity gate working with weak cross-Kerr nonlinearity. By the photon number resolving detection on one of the output quantum bus or communication beams, the setup will realize a near deterministic transformation to a Bell state for every entangling attempt. With the simple threshold detectors, on the other hand, the system can still reach a considerable success probability of $50\%$ per try. The decoherence effect caused by photon absorption losses in the operation is also discussed.
\end{abstract}

\newpage

\maketitle

\section{Introduction}\label{section1}

Entangled photon pair sources have wide applications in quantum cryptography, teleportation, dense coding, and quantum computation because of the simplicity to manipulate such bi-photon states with the current technologies \cite{p1,p2}. The efficient generation of these entangled photon pairs therefore attracts extensive research in the recent years. At the present time the primary method of experimentally producing entangled photon pairs is the parametric down-conversion (PDC) in $\chi^{(2)}$ crystals \cite{optics1, optics2} (here we only cite the seminal works). A PDC source emits entangled photon pairs with their numbers following the approximate Poisson distribution in the weak down-conversion regime \cite{k-b}. In addition to the polarization modes, PDC pairs are correlated in spatial modes, wave lengths, etc., and these extra correlations might limit the flexibility of such pair source. An ideal entangled pair source should output the following maximally entangled states 
\begin{eqnarray}
|\Phi^{\pm}\rangle&=& \frac{1}{\sqrt{2}}(|HH\rangle_{A,B}\pm|VV\rangle_{A,B}),\nonumber\\
|\Psi^{\pm}\rangle&=& \frac{1}{\sqrt{2}}(|HV\rangle_{A,B}\pm|VH\rangle_{A,B}),
\label{1}
\end{eqnarray}
which are correlated only in their polarization modes $H$ and $V$ representing the horizontal and vertical polarization, respectively (the indexes $A$ and $B$ denote the sharing parties of the entanglement). The generation of these discrete entangled state, which are called the Bell states, is important to various practical applications.
In reality, due to the possible decoherence and imperfection, the generated photon pairs could largely deviate from the Bell states, and the performance of the quantum information protocols based on them would be degraded. To improve on the quality of the entangled pairs, one will demand the procedures of entanglement concentration \cite{ben1} and entanglement purification \cite{ben2} to make the generated states be closer to the Bell states. These operations would be difficult to perform if the states of the generated entangled pairs from a PDC or other source are unknown. 

In this work we propose a design for the setup that transforms an input photon pair 
in arbitrary rank-four mixed state,
\begin{eqnarray}
\rho_{in}&=&\sigma_1|\Lambda_1\rangle\langle\Lambda_1|+\sigma_2|\Lambda_2%
\rangle\langle\Lambda_2| 
+\sigma_3|\Lambda_3\rangle\langle\Lambda_3|
+\sigma_4|\Lambda_4\rangle
\langle\Lambda_4|,
\label{2}
\end{eqnarray}
to an maximally entangled states in Eq. (\ref{1}). 
In the above equation $\sigma_i$ are the eigenvalues of $\rho_{in}$ and $|\Lambda_i\rangle$, the pure state components as its eigenvectors, are the linear combinations of $|\Phi^{\pm}\rangle$ and $|\Psi^{\pm}\rangle$. 
There are dual functions of this setup. If $\rho_{in}$ is the tensor product of the states of two independent single photons, it will naturally work as an entangler. Single photon sources have been close to practical applications recently (for an overview on the single photon sources based on parametric down-conversion in $\chi^{(2)}$ materials, see \cite{single-photon}). A bright source of heralded narrow-band single photons has been experimentally realized using a double-resonant optical parametric oscillator (OPO) \cite{opo}. Also with electromagnetic induced transparency (EIT) technique, it is possible to generate single photons with narrow bandwidth which is actually the same as that of the control laser (typically about 1 MHz). If we entangle the single photons from these sources only in their polarization modes, the entangled pairs with narrow bandwidth will be extremely suitable for long-distance quantum communications. In the system we transmit the correlation between the polarization modes of two individual photons by two identical laser beams in coherent state. These so-called quantum bus (qubus) or communication beams interact with the single photons through cross-phase modulation (XPM) processes in Kerr media. The state of the qubus beams in the transmission channel is entangled coherent state or coherent-state superposition known as cat state \cite{g-v}. The decoherence effects on the superpositions of coherent states being transmitted through a standard optical fiber are dominated by the channel loss due to photon absorption (see, e.g., \cite{n-c}). It will result in a trade-off between the fidelity of an entangled photon pair and the success probability of the entangling operation. However, such trade-off is controllable by adjusting the parameters of the system so that the entangled photon pairs of high fidelity could be generated. The second function of the setup is the purification of noisy photon pairs. XPM processes have been proposed for the realization of entanglement concentration and purification \cite{sd1,sd2}. Different from all previous schemes, our setup could effectively perform as a purifier for photon pairs without the knowledge about their exact states ($\sigma_i$ and $|\Lambda_i\rangle$ in Eq. (\ref{2}) could be unknown).

The rest of the paper is organized as follows. In Sec. II we illustrate the linear optical circuits and quantum non-demolition detections to project the pure state components $|\Lambda_i\rangle$ in Eq. (\ref{2}) to the same superposition of two Bell states. The weak nonlinearity based parity gate to realize the Bell states is described in Sec. III. The decoherence effects caused by photon absorption losses and the optimum performance of the setup in the decoherence environment are discussed in Sec. IV. The main results of the work are summarized in the last section.

\begin{figure}
\includegraphics[width=128truemm]{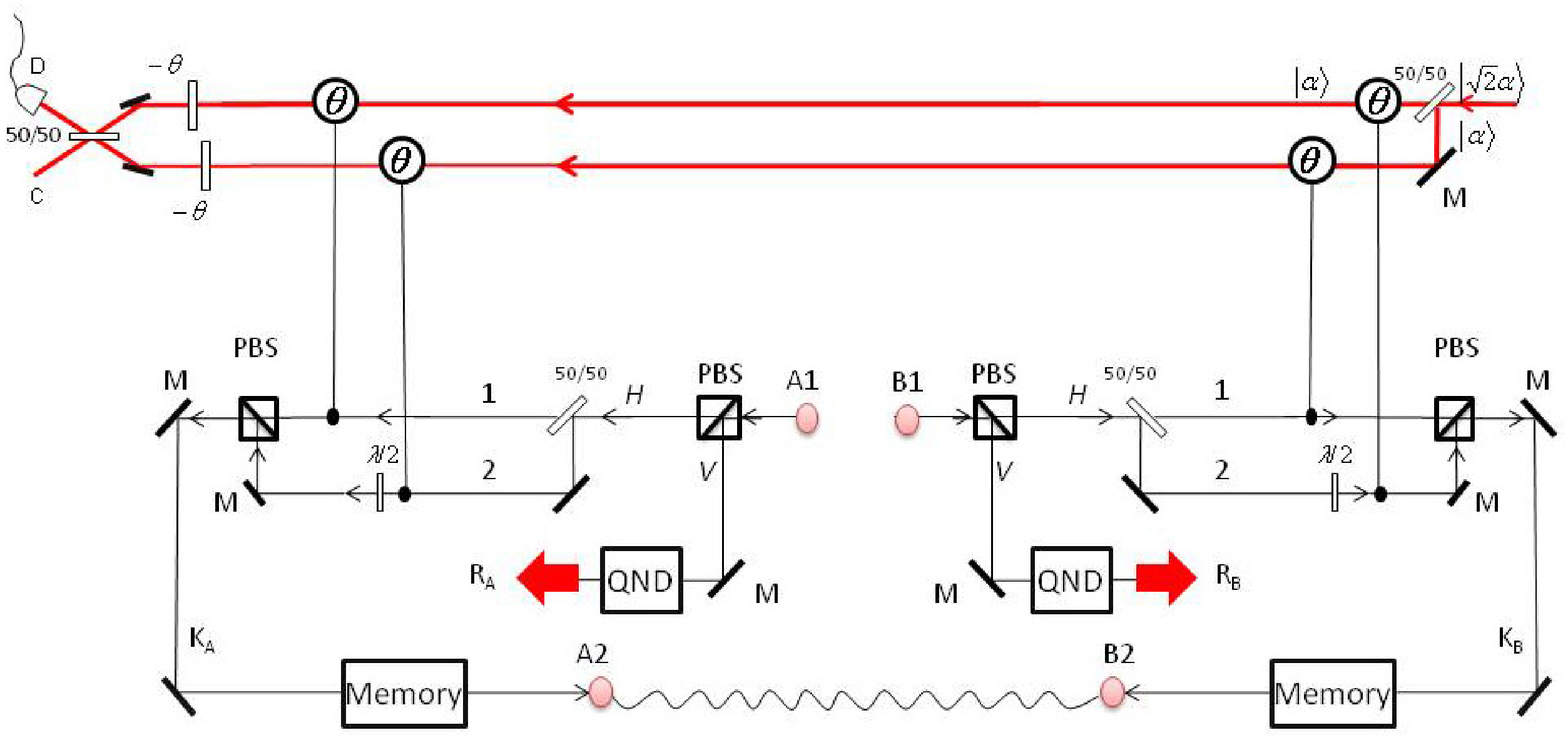}
\caption{Schematic diagram of the setup to maximally entangle a photon pair in arbitrary rank-four mixed state. A photon pair input from the terminals A1 and B1 is transformed to a which-path space by two PBS. One coherent beam in the QND modules shown in Fig. 2 is coupled to $V$ component only, and their response patterns project out a superpositon of two Bell states over one of the output port pairs  $\{K_A, K_B\}$, $\{K_A, R_B\}$, $\{R_A, K_B\}$ and $\{R_A, R_B\}$. The 50/50 beam splitters transform the photon pair state to a doubled dimensional space, so that the qubus beams will entangle the 
projected-out components through the XPM processes in Kerr media. The half-wave plates ($\lambda/2$) and the other PBS covert the photon pair state back to polarization space. The detection of the photon detector D heralds the generation of a Bell state over A2 and B2.   } 
\end{figure}

\section{Decomposition of input bi-photon state} \label{section2}

To realize a maximally entangled state out of a photon pair in arbitrary rank-four mixed state of Eq. (\ref{2}), we first process the individual photons as shown in Fig. 1. The pure state components in the mixed state of a photon pair are represented in the separable basis $\{|HH\rangle, |HV\rangle, |VH\rangle, |VV\rangle\}$ as 
\begin{eqnarray}
|\Lambda_i\rangle=c^i_1|HH\rangle_{A,B}+c^i_2|HV\rangle_{A,B}+c^i_3|VH\rangle_{A,B}+c^i_4|VV\rangle_{A,B},
\label{3}
\end{eqnarray}
for $i=1,\cdots,4$. By two polarization beam splitters (PBS), a photon pair sent into the terminals A1-A2 in Fig. 1 is converted to a which-path space for further processing. Then, with two 50/50 beam splitters implementing the following $U(2)$ transformation
\begin{eqnarray}
\hat{c}^{\dagger}_{in}&\rightarrow & \frac{1}{\sqrt{2}}(\hat{c}^{\dagger}_{out}+\hat{d}^{\dagger}_{out})\nonumber\\
\hat{d}^{\dagger}_{in}&\rightarrow &\frac{1}{\sqrt{2}}(\hat{c}^{\dagger}_{out}-\hat{d}^{\dagger}_{out})
\label{4}
\end{eqnarray}
on both side A and B, we extend the $2\times 2$-dimensional input to a space of $4\times 4$ dimension:

\begin{eqnarray}
|\Lambda_i\rangle &\rightarrow &\frac{c^i_1}{2}(\hat{a}^{\dagger}_{H,1}+\hat{a}^{\dagger}_{H,2})(\hat{b}^{\dagger}_{H,1}+\hat{b}^{\dagger}_{H,2})|0\rangle+\frac{c^i_2}{2} (\hat{a}^{\dagger}_{H,1}+\hat{a}^{\dagger}_{H,2})(\hat{b}^{\dagger}_{V,3}+\hat{b}^{\dagger}_{H,4})|0\rangle\nonumber\\
&+& \frac{c^i_3}{2}(\hat{a}^{\dagger}_{V,3}+\hat{a}^{\dagger}_{V,4})(\hat{b}^{\dagger}_{H,1}+\hat{b}^{\dagger}_{H,2})|0\rangle+\frac{c^i_4}{2} (\hat{a}^{\dagger}_{V,3}+\hat{a}^{\dagger}_{V,4})(\hat{b}^{\dagger}_{V,3}+\hat{b}^{\dagger}_{H,4})|0\rangle,
\label{5}
\end{eqnarray}
where $\hat{a}^{\dagger}$ represents the modes on side A and $\hat{b}^{\dagger}$ the modes on side B. The path $3$ and $4$ at two locations are not shown in Fig. 1. Two half-wave plates ($\lambda/2$ in Fig. 1) on path $2$ and $4$ and a PBS continue to transform the state to the superposition of four product states:
\begin{eqnarray}
&&\frac{c^i_1}{2}(|H\rangle_{K_A}+|V\rangle_{K_A})(|H\rangle_{K_B}+|V\rangle_{K_B})+\frac{c^i_2}{2}(|H\rangle_{K_A}+|V\rangle_{K_A})(|H\rangle_{R_B}+|V\rangle_{R_B}) \nonumber\\
&+& \frac{c^i_3}{2}(|H\rangle_{R_A}+|V\rangle_{R_A})(|H\rangle_{K_B}+|V\rangle_{K_B})+\frac{c^i_4}{2}(|H\rangle_{R_A}+|V\rangle_{R_A})(|H\rangle_{R_B}+|V\rangle_{R_B}).
\label{6}
\end{eqnarray}
If we project out the piece from any one of the ports $\{K_A, K_B\}$, $\{K_A, R_B\}$, $\{R_A, K_B\}$ and $\{R_A, R_B\}$, the total output from all $|\Lambda_i\rangle$ will be proportional to $|\Phi^+\rangle+|\Psi^+\rangle$, which can be separated by a parity gate to obtain a Bell state.

\begin{figure}
\includegraphics[width=86truemm]{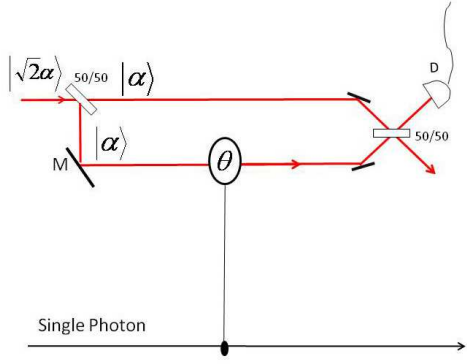}
\caption{Schematic diagram of QND module. The interaction
of one of the coherent beams $|\alpha\rangle$ with a single photon state in Kerr medium generates an extra phase $\theta $
on the coherent beam. The 50/50 beam splitter transforms two coherent states 
$|\alpha \rangle _{1}|\beta\rangle_2$, where $|\beta\rangle
_{2}=|\alpha \rangle _{2}$ or $|\alpha e^{i\theta
}\rangle _{2}$, to $|\frac{\alpha -\beta }{\sqrt{2}}\rangle _{1}|
\frac{\alpha +\beta }{\sqrt{2}}\rangle _{2}$. A response of the threshold D 
indicates that the two beams $|\alpha \rangle _{1}$ and $|\beta \rangle _{2}$ are
different with $|\frac{\alpha-\beta }{\sqrt{2}}\rangle _{1}\neq
|0\rangle _{1}$.  } 
\end{figure}

We here apply the quantum non-demolition (QND) modules illustrated in Fig. 2 to implement the projection. In a module one of the beams in coherent state $|\alpha\rangle$ interacts with a single photon through an XPM process in Kerr medium ($\chi^{(3)}$ medium), which is described by the effective Hamiltonian ${\cal H}=-\hbar \chi \hat{n}_i\hat{n}_j$ ($\chi $, the nonlinear intensity, is proportional to the third order optical nonlinearity $\chi^{(3)}$, and $\hat{n}_{i/j}$ the number operator of the coupling modes). Then the coherent beam is compared with another untouched one by means of a 50/50 beam splitter and a photon number non-resolving detector. Without the loss of generality, we let the coherent beams in two QND modules be coupled to $V$ component on each side as in Fig. 1. This will effect the transformation to the following state from Eq. (\ref{6}): 
\begin{eqnarray}
&&\frac{c^i_1}{2}(|H\rangle_{K_A}+|V\rangle_{K_A})(|H\rangle_{K_B}+|V\rangle_{K_B})|0\rangle_{A,1}|\sqrt{2}\alpha\rangle_{A,2} |0\rangle_{B,1}|\sqrt{2}\alpha_0\rangle_{B,2}\nonumber\\
&+&\frac{c^i_2}{2}(|H\rangle_{K_A}+|V\rangle_{K_A})(|H\rangle_{R_B}+|V\rangle_{R_B})|0\rangle_{A,1}|\sqrt{2}\alpha\rangle_{A,2} 
|\frac{\alpha e^{i\theta}-\alpha}{\sqrt{2}}\rangle_{B,1}|\frac{\alpha e^{i\theta}+\alpha}{\sqrt{2}}\rangle_{B,2} \nonumber\\
&+& \frac{c^i_3}{2}(|H\rangle_{R_A}+|V\rangle_{R_A})(|H\rangle_{K_B}+|V\rangle_{K_B})|\frac{\alpha e^{i\theta}-\alpha}{\sqrt{2}}\rangle_{A,1}|\frac{\alpha e^{i\theta}+\alpha}{\sqrt{2}}\rangle_{A,2}|0\rangle_{B,1}|\sqrt{2}\alpha\rangle_{B,2} \nonumber\\
&+& \frac{c^i_4}{2}(|H\rangle_{R_A}+|V\rangle_{R_A})(|H\rangle_{R_B}+|V\rangle_{R_B})|\frac{\alpha e^{i\theta}-\alpha}{\sqrt{2}}\rangle_{A,1}|\frac{\alpha e^{i\theta}+\alpha}{\sqrt{2}}\rangle_{A,2}|\frac{\alpha e^{i\theta}-\alpha}{\sqrt{2}}\rangle_{B,1}|\frac{\alpha e^{i\theta}+\alpha}{\sqrt{2}}\rangle_{B,2} .
\label{7}
\end{eqnarray}
Now we could project out the proper bi-photon components by detecting the local coherent states carrying the indexes $A,1$ and $B,1$ in Eq. (\ref{7}).
For a sufficiently large $|\alpha|$, the amplitude of the state $|\frac{\alpha e^{i\theta}-\alpha}{\sqrt{2}}\rangle$ could be large enough as well, and it will be possible to certainly discriminate it from $|0\rangle$ using simple photon number non-resolving detector described by the positive-operator-value measure (POVM) elements \cite{k-b}
\begin{eqnarray}
\Pi_0&=&\sum_{n=0}^{\infty}(1-\eta_D)^n|n\rangle\langle n|\nonumber\\
\Pi_1&=&I-\Pi_0=\sum_{n=0}^{\infty}\{1-(1-\eta_D)^n\}|n\rangle\langle n|,
\label{povm}
\end{eqnarray}
where $\eta_D$ is the photon detection efficiency.
Such detector can be an avalanche photodiode (APD) which outputs the same signal no matter how many photons are captured. The on-off reaction patterns of the APDs in two QND modules, which are summarized in Tab. \ref{tb1}, therefore project each pure state component in Eq. (\ref{2}) to the 
same Bell state superposition, $|\Phi^+\rangle+|\Psi^+\rangle$, over four pairs of output ports. As seen from Eq. (\ref{7}), the projection result is relevant only to the output ports no matter what coefficents $c^i_k$ should be. Once the output ports for the projected-out state are heralded by the QND detections, the output will the same from any input state even if its $\sigma_i$ and $|\Lambda_i\rangle$ in Eq. (\ref{2}) are unknown in operation.

\begin{table}
\begin{tabular}{|c|c|}\hline
 Module A---Module B & Projected-out State\\ \hline
off---off  &  $|\Phi^+\rangle_{K_A,K_B}+|\Psi^+\rangle_{K_A,K_B}$\\ \hline
on---on &   $|\Phi^+\rangle_{R_A,R_B}+|\Psi^+\rangle_{R_A,R_B}$ \\ \hline
off---on &     $|\Phi^+\rangle_{K_A,R_B}+|\Psi^+\rangle_{K_A,R_B}$    \\ \hline
on---off &    $|\Phi^+\rangle_{R_A,K_B}+|\Psi^+\rangle_{R_A,K_B}$   \\ \hline
\end{tabular}
\caption{The response patterns of two QND modules on side A, B and the corresponding intermediate states generated from the projection by the QND modules. These patterns are for the design in Fig. 1. If we use one more QND module on each side, the patterns to herald the generation of the intermediate state will be which pair of QND modules are on.}
\label{tb1}
\end{table}

An alternative design is to decompose the input state in Eq. (\ref{2}) with respect to the basis $\{|++\rangle, |+-\rangle, |-+\rangle, |--\rangle\}$, where $|\pm\rangle=1/\sqrt{2}(|H\rangle\pm |V\rangle)$. Now, as shown in Fig. 3, the two PBS to convert the input from the polarization space to the which-path space in Fig. 1 should be replaced by two PBS$\pm$, which transmit $|+\rangle$ component and reflect $|-\rangle$ component. Then, with two other PBS on both sides, we will still get the superposition of two Bell states over the pairs of output ports. The states projected out of all $|\Lambda_i\rangle$ over port $K_A$ and $K_B$ will be $|\Phi^+\rangle+|\Psi^+\rangle$, but it will be $|\Phi^+\rangle-|\Psi^+\rangle$ over port $R_A$ and $R_B$, and $|\Phi^-\rangle-|\Psi^-\rangle$ over $K_A$, $R_B$ and $R_A$, $K_B$.

\begin{figure}
\includegraphics[width=90truemm]{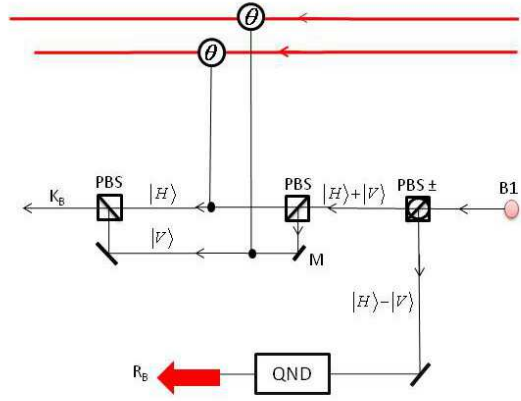}
\caption{Schematic diagram of an alternative design to the setup in Fig. 1. Here the input photon pair states are decomposed by PBS$\pm$ in the space spanned by $\{|++\rangle, |+-\rangle, |-+\rangle, |--\rangle\}$. The PBS further separate $H$ and $V$ components for the interaction with the qubus beams. By this design the projected-out separable states are different over the four pairs of output ports.} 
\end{figure}

\section{Double XPM parity gate} \label{section3}

The states $|\Phi^{\pm}\rangle$ are conventionally called the even parity Bell states, and $|\Psi^{\pm}\rangle$ the odd parity Bell states. A parity gate is used to separate these different parity components. Bi-photon parity gates could be built with weak cross-Kerr nonlinearity plus homodyne detection \cite{cc1} or photon number resolving detection \cite{cc2, kerr}. Here we present a double XPM parity gate based on coherent states comparison \cite{andersson}. Its advantages will be demonstrated in a decoherence environment discussed in the next section.  

As illustrated in Fig. 1, we first interact two qubus or communication beams in the identical coherent state $|\alpha\rangle$ with photon B through two Kerr nonlinearities, which effect the unitary operations $U_{K,i}=exp(i\chi t\hat{n}_{H/V} \hat{n}_{i})$ ($i=1,2$ and $t$ the interaction time). Here the coupling of the qubus beams to the modes going to $K_B$ or $R_B$ is determined by the classically feed-forwarded measurement results of the QND module. Overall, such XPM processes evolve the state of the whole system as follows: 
\begin{eqnarray}
U_{K,1}U_{K,2}\{(|\Phi^+\rangle+|\Psi^+\rangle)|\alpha\rangle_1|\alpha\rangle_2\}
&=&(|H\rangle_A+|V\rangle_A) \{ U_{K,1}|\alpha \rangle_1 ~U_{K,2}(|H\rangle_B|\alpha\rangle_2)  +  U_{K,1}(|V\rangle_B |\alpha \rangle_1) 
~U_{K,2}|\alpha\rangle_2 \}\nonumber\\
&=& (|H\rangle_A+|V\rangle_A)(|H\rangle_B|\alpha \rangle_1|\alpha e^{i\theta}\rangle_2 
+ |V\rangle_B|\alpha e^{i\theta}\rangle_1 |\alpha \rangle_2) =|\Psi_1\rangle_{tot},
\label{UK}
\end{eqnarray}
where $\theta=\chi t$, and the global coefficient is neglected. At location B, the first qubus beam is coupled to the single photon mode on track 2 while the second to that on track 1. After the XPM processes, photon B will be stored in any type of quantum memory and the qubus beams will be sent to location A, where we use Kerr nonlinearities inducing the same phase shift to couple them with photon A in a different way---the first beam will be coupled to the mode on track 1 and the second to that on track 2. With which single photon modes (going to $K_A$ or $R_A$) the qubus beams should be coupled to is also decided by the QND module's detection at location A. Then, two phase shifters of $-\theta$ are applied on the qubus beams to transform the state of the system to 
\begin{eqnarray}
|\Psi_2\rangle_{tot}=(|HH\rangle+|VV\rangle)|\alpha\rangle_1|\alpha\rangle_2 
+|HV\rangle|\alpha e^{i\theta}\rangle_1|\alpha e^{-i\theta}\rangle_2+|VH\rangle |\alpha e^{-i\theta}\rangle_1|\alpha e^{i\theta}\rangle_2. 
\end{eqnarray}
Like the process in a QND module in Fig. 2, the comparison of the two qubus beam states in the above equation is performed by a 50/50
beam splitter and a photon detector D. The final state before the photon detection is
\begin{eqnarray}
|\Psi_3\rangle_{tot}&=&(|HH\rangle+|VV\rangle)|0\rangle_1|\sqrt{2}\alpha \rangle_2\nonumber\\
&+&|HV\rangle |i\sqrt{2}~\alpha \sin\theta\rangle_1~|\sqrt{2}%
~\alpha\cos\theta\rangle_2
+|VH\rangle~|-i\sqrt{2}~\alpha \sin\theta\rangle_1~|\sqrt{2}%
~\alpha\cos\theta\rangle_2.
\label{parity}
\end{eqnarray}

In the Fock state basis the first coherent state with the odd parity bi-photon component (the linear combination of $|HV\rangle$ and $|VH\rangle$) 
is
\begin{eqnarray}
|\pm i\sqrt{2}~\alpha \sin\theta\rangle_1=e^{-(\alpha\sin\theta)^2}\sum_{n=0}^{\infty} \frac{(\pm i\sqrt{2}~\alpha \sin\theta)^n}{\sqrt{n!}}|n\rangle_1.
\end{eqnarray}
A Bell state can be now obtained by the Fock state projectors $\Pi_i=|i\rangle\langle i|$ ($i=0,\cdots,\infty$) on this qubus beam. 
If the measured value of the photon number is $n=0$, the resulting Bell state will be $|\Phi^+\rangle$; 
if the photon number $n\neq0$, we will obtain the state $\frac{1}{\sqrt{2}}(e^{i\frac{\pi}{2}n}|HV\rangle+e^{-i\frac{\pi}{2}n}|VH\rangle)$, which could be either $|\Psi^{+}\rangle$ or $|\Psi^{-}\rangle$ with the respective probability $P_n=e^{-2(\alpha\sin\theta)^2}\frac{(\sqrt{2}\alpha\sin\theta)^{2n}}{n!}$. 
The error in the operation is due to the overlap between $|0\rangle$ and $|\pm i\sqrt{2}~\alpha \sin\theta\rangle$, which gives the error probability $P_{err}=e^{-2(\alpha\sin\theta)^2}$. If $|\alpha\sin\theta|$ is large enough, each entangling operation will be almost deterministic.

In practical implementation, the Fock state projections $\Pi_i=|i\rangle\langle i|$ can be performed by using an extra QND module in Fig. 2 \cite{dd1,dd2}.
Now the coherent beams in the module are initially in the state $|\gamma\rangle$. One of the coherent beams will be coupled to $|\pm i\sqrt{2}~\alpha \sin\theta \rangle$ through a cross Kerr medium, effecting the following transformation
\begin{eqnarray}
|\pm i\sqrt{2}~\alpha \sin\theta \rangle |\gamma\rangle 
|\gamma\rangle \rightarrow  e^{-(\alpha \sin\theta)^2}\sum_{n=0}^{\infty}\frac{(\pm i\sqrt{2}~\alpha \sin\theta)^{n}}{\sqrt{n!}} |n \rangle |\gamma e^{in\theta}\rangle|\gamma\rangle.
\label{pro}
\end{eqnarray}
After the 50/50 beam splitter in the QND module, the coherent state of the two detecting beams will be $\left\vert
\frac{\gamma e^{in\theta}-\gamma}{\sqrt{2}}\right\rangle \left\vert
\frac{\gamma e^{in\theta}+\gamma}{\sqrt{2}}\right\rangle $, where $n$ is from $0$ to infinity. Since each $\left\vert
\frac{\gamma e^{ik\theta}-\gamma}{\sqrt{2}}\right\rangle$ has a certain Poisson distribution of photon numbers from $n_k$ to $n_k'$, the number non-resolving detector's POVM element $\Pi_1$ in Eq. (\ref{povm}) actually functions as $\Pi_{1,k}=\sum_{m=n_k}^{n_k'}(1-(1-\eta)^m)|m\rangle\langle m|$ on each of these coherent state components. The photon number non-resolving detector in this QND module should be a sensor outputting the signals (voltage or current) proportional to the total photon detection probabilities $\langle \frac{\gamma e^{ik\theta}-\gamma}{\sqrt{2}}|\Pi_{1,k}|\frac{\gamma e^{ik\theta}-\gamma}{\sqrt{2}}\rangle$. Given a large coherent beam amplitude $|\gamma|$, the output signals for the different $\left\vert
\frac{\gamma e^{ik\theta}-\gamma}{\sqrt{2}}\right\rangle$ can be mutually distinct, and the photon number resolving detection will be implemented in such indirect way.

On the other hand, if we are only provided with APDs, the threshold detectors registering photon or no photon, the total success probability 
will be lowered to $50\%$ per try. We could also compare the second beam ($C$ in Fig. 1), which is either in the state $|\sqrt{2}\alpha\rangle_2$ 
or in the state $|\sqrt{2}~\alpha\cos\theta\rangle_2$ from Eq. (\ref{parity}), with another prepared state $|\sqrt{2}~\alpha\cos\theta\rangle_3$ at location A. Then we have such detection pattern for the generated states: if the second coherent beam is responded by a detector, we will realize the Bell state $|\Phi^{+}\rangle$ with the success probability $50\%$; if the first coherent beam is responded, on the other hand, what we obtain will be an unknown superposition of $|\Psi^{\pm}\rangle$ by the other $50\%$ probability. 

The entangled photon pair generation rate of the system could be limited by the dead time of the photon detector in the QND modules, but the detection inefficiency as in single photons detection can be overcome by using sufficiently bright qubus beams. Since the photons of the generated pairs are only correlated in their polarizations, the wavelengths and bandwidths of the entangled pairs are those of the single photon sources.

\section{Decoherence effects} \label{section4}
In the above discussions, we assume that the losses of the qubus beams are negligible. In reality, however, there could be various losses, especially when the two qubus beams are sent through a long-distance optical fiber to another location.
Another source of decoherence is the instability of the phase difference of two qubus beams due to the fluctuation of their transmission times through the optical fibers. Such phase noise can be eliminated by sending them through the same fiber within a short time interval, as has been demonstrated in a recent experiment for single photon signals \cite{trans}.
In this section we will focus on the decoherence due to losses, and present the optimum performance of the system in such decoherence environment.

\subsection{Realistic cross-Kerr nonlinearity}

First, we look at the Kerr media for implementing the required XPM processes. A normal Kerr crystal could be too weak to generate a sufficient phase shift \cite{kerr1}. The satisfactory performance of a parity gate working with such cross-Kerr nonlinear crystal could also be spoiled by phase noise \cite{kerr}. Beyond the normal Kerr nonlinearities, the possible candidates for realizing the necessary XPM process are the atomic systems working under EIT conditions \cite{eit}, where the phase noise model in \cite{kerr} may not be applicable. Although the dissipative losses of the photonic modes in an EIT medium could also exist, we only need to generate a very small phase shift $\theta\ll 1$ in Eq. (\ref{UK}), and it intrinsically lowers the losses of single photon and coherent beam \cite{2005}. By using a sufficiently bright coherent beams $|\alpha\rangle$ we can still realize the high efficiency of the entangler. This is one of the advantages of our setup.

In \cite{bhe} some of us studied the decoherence due to the losses in the XPM processes for generating cat states as in Eq. (\ref{UK}). It is shown 
 that the good coherence of the generated state in Eq. (\ref{UK}) and the reasonable qubus beam intensity $|\alpha|^2$ are possible with the appropriate signal field detuning and interaction time. 

\subsection{Channel loss}

Since the phase noise of two qubus beams can be eliminated if they are transmitted as the reference of each other, the decoherence effect on the total state of the system is predominantly from their losses in optical fiber. The decoherence caused by the losses in the XPM processes could be also included in this part by extending the effective length of the optical fiber. There have been studies on such decoherence in hybrid or qubus quantum repeater protocols \cite{van-loock, ladd, munro, h-repeater}. By our design the initial qubus state is either an entangled coherent state as in Eq. (\ref{UK}) or the product of a cat state and a coherent state,
\begin{eqnarray}
(|H\rangle_B|\frac{\alpha-\alpha e^{i\theta}}{\sqrt{2}} \rangle_1 
+ |V\rangle_B|\frac{\alpha e^{i\theta}-\alpha}{\sqrt{2}}\rangle_1) |\frac{\alpha+\alpha e^{i\theta}}{\sqrt{2}}\rangle_2,
\label{cat}
\end{eqnarray}
by applying a transformation with a 50/50 beam splitter. The decoherence on a cat state due to loss can be found in \cite{g-v}, and that for the state in Eq. (\ref{cat}) is also studied in \cite{munro}. Here we discuss the decoherence of the entangled coherent state in an amplitude-damping channel by solving the master equation involving the two qubus beams and the single photon qubit at location B  
\begin{eqnarray}
\frac{d\rho}{dt}=\frac{\gamma}{2}\sum_{i=1}^2\{[\hat{a}_i\rho,\hat{a}_i^{\dagger}]+[\hat{a}_i,\rho\hat{a}_i^{\dagger}]\}=\hat{{\cal D}}\rho,
\label{loss}
\end{eqnarray}
where $\gamma$ is the loss rate of a coherent beam in optical fiber, the sum is over two qubus modes, and the initial state is $\rho(t_0)=|\Psi_0\rangle\langle\Psi_0|$, with $|\Psi_0\rangle=|H\rangle_B|\alpha \rangle_1|\alpha e^{i\theta}\rangle_2 
+ |V\rangle_B|\alpha e^{i\theta}\rangle_1 |\alpha \rangle_2$. During a period of time $t-t_0$ for the qubus beams being sent to location A, the initial pure state will be decohered to 
\begin{eqnarray}
\rho(t)=exp\{\hat{{\cal D}}(t-t_0)\}\rho(t_0)=\frac{1+|\xi|^2}{2}|\Phi^1\rangle_{B,1,2}\langle\Phi^1|+\frac{1-|\xi|^2}{2}|\Phi^2\rangle_{B,1,2}\langle\Phi^2|
\label{GG}
\end{eqnarray}
where
\begin{eqnarray}
|\Phi^1\rangle_{B,1,2}&=&|H\rangle_B|\sqrt{\eta}\alpha \rangle_1|\sqrt{\eta}\alpha e^{i\theta}\rangle_{2}
+|V\rangle_B|\sqrt{\eta}\alpha e^{i\theta}\rangle_1|\sqrt{\eta}\alpha \rangle_{2},\nonumber\\
|\Phi^2\rangle_{B,1,2}&=&|H\rangle_B|\sqrt{\eta}\alpha \rangle_1\sqrt{\eta}\alpha e^{i\theta}\rangle_{2}
-|V\rangle_B|\sqrt{\eta}\alpha e^{i\theta}\rangle_1|\sqrt{\eta}\alpha \rangle_{2},
\label{c}
\end{eqnarray}
and
\begin{eqnarray}
|\xi|^2=|\langle\sqrt{1-\eta}\alpha e^{i\theta}|\sqrt{1-\eta}\alpha\rangle|^2=e^{-(1-\eta)|\alpha e^{i\theta}-\alpha|^2},~~~~~\eta=e^{-\gamma(t-t_0)}. 
\label{fidelity}
\end{eqnarray}
As the result, the fidelity of the whole state after the qubus beams going through the transmission channel will be lowered to $F=(1+|\xi|^2)/2$.

\subsection{Optimum efficiency-fidelity relation}
Through the non-local entangling operation of the parity gate discussed in the last section, what we realize in a decoherence environment will be a mixed state 
\begin{eqnarray}
\rho_{out}=F|\Psi^1\rangle\langle\Psi^1|+(1-F)|\Psi^2\rangle\langle\Psi^2|,
\label{out}
\end{eqnarray}
where
\begin{eqnarray}
|\Psi^1\rangle&=&\frac{1}{\sqrt{2}}(|HH\rangle+|VV\rangle)|0\rangle_1|\sqrt{2 \eta}\alpha \rangle_2\nonumber\\
&+&\frac{1}{\sqrt{2}}\left(|HV\rangle |i\sqrt{2 \eta}~\alpha \sin\theta\rangle_1+|VH\rangle~|-i\sqrt{2 \eta}~\alpha \sin\theta\rangle_1\right )|\sqrt{2 \eta}
~\alpha\cos\theta\rangle_2,
\label{d1}
\end{eqnarray}
\begin{eqnarray}
|\Psi^2\rangle&=&\frac{1}{\sqrt{2}}(|HH\rangle-|VV\rangle)|0\rangle_1|\sqrt{2 \eta}\alpha \rangle_2\nonumber\\
&-&\frac{1}{\sqrt{2}}\left(|HV\rangle |i\sqrt{2 \eta}~\alpha \sin\theta\rangle_1-|VH\rangle~|-i\sqrt{2 \eta}~\alpha \sin\theta\rangle_1\right)
|\sqrt{2 \eta}~\alpha\cos\theta\rangle_2,
\label{d2}
\end{eqnarray}
and $F=(1+e^{-(1-\eta)|\alpha e^{i\theta}-\alpha|^2})/2$.
From such output we will obtain a mixture of two Bell states of the same parity.
The error probability in this case is that of the vacuum component in $|\pm i\sqrt{2 \eta}~\alpha \sin\theta\rangle$. Here we assume $\theta\ll 1$, and the relation between the the fidelity and the success probability is then given as
\begin{eqnarray}
P=1-e^{-2 \eta(\alpha \sin\theta)^2}=1-(2F-1)^{\frac{2\eta}{1-\eta}}.
\label{bound}
\end{eqnarray}
Fig. 4 demonstrates this upperbound efficiency-fidelity relation for some different transmission coefficients $\eta$. 
The parity gate based on coherent states comparison therefore outperforms the other schemes \cite{van-loock, ladd, munro, h-repeater} in the decoherence environment of lossy optical fiber. To eliminate the unwanted component $|\Psi^2\rangle$ in Eq. (\ref{out}), we can choose the proper parameters such that $|\xi|^2\sim 1$. For example, if $|\alpha|$ is set to the order of $10^3$ and $\theta$ is in the order of $10^{-5}$, the fidelity of the state in Eq. (\ref{out}) will be larger than $1-10^{-4}$. On the other hand, if we want to achieve both high efficiency and fidelity, the qubus beam loss in the transmission channel should be sufficiently small. Demanding $F=0.99$ and $P=0.99$, for instance, we will need a good transmission coefficient $\eta\sim 0.99$. 

\begin{figure}
\includegraphics[width=87truemm]{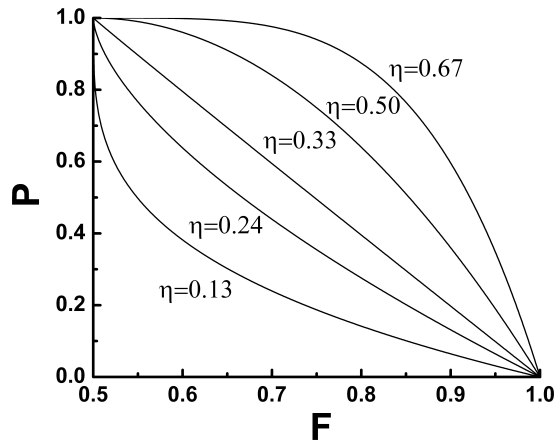}
\caption{The upperbound efficiency-fidelity relation of Eq. (\ref{bound}) for the performance of the system under the qubus beam losses. The curves from the upper to the lower correspond to the transmission $\eta=0.67$, $0.50$, $0.33$, $0.24$ and $0.13$ of the optical fiber, respectively. } 
\end{figure}

We could also obtain an approximate Bell state simply by comparing the state of two output qubus beams with another tensor product state $|0\rangle_3\sqrt{2\eta}\alpha\rangle_4$. The circuit for the purpose consists of two 50/50 beam splitters and two photon detectors. 
Following the notation in \cite{coherent-circuit}, we represent the transformation of the linear optical circuit as ($\alpha$ is assumed to be real)
\begin{eqnarray}  
\left(\begin{array}{c}\gamma_1\\
 \gamma_2\\
 \gamma_3\\
 \gamma_4\\
 \end{array}\right)=\left(
\begin{array}{cccc}
\frac{1}{\sqrt{2}} & 0 & -\frac{1}{\sqrt{2}} & 0 \\ 
0& \frac{1}{\sqrt{2}} & 0 & -\frac{1}{\sqrt{2}}\\
\frac{1}{\sqrt{2}}& 0 & \frac{1}{\sqrt{2}} & 0\\
0 & \frac{1}{\sqrt{2}} & 0 & \frac{1}{\sqrt{2}}\\
\end{array}%
\right)\left(\begin{array}{c}\beta_1\\
 \beta_2\\
 0\\
\sqrt{2\eta}\alpha \\
 \end{array}\right).
\end{eqnarray}
The state $|0\rangle_3\sqrt{2\eta}\alpha\rangle_4$ for comparison is input from the third and fourth input port, and $|\beta_1\rangle_1|\beta_2\rangle_2$ is either $|0\rangle_1|\sqrt{2\eta}\alpha\rangle_2$ or $|\pm i\sqrt{2 \eta}~\alpha \sin\theta\rangle_1|\sqrt{2 \eta}~\alpha\cos\theta\rangle_2$ from Eqs. (\ref{d1}) and (\ref{d2}). If the first or the second beam of the output state, $|\gamma_1\rangle_{1}|\gamma_2\rangle_{2}|\gamma_3\rangle_{3}|\gamma_4\rangle_{4}$, is detected, we will realize an approximate Bell state $|\Psi^{-}\rangle$. Under the condition $|\alpha \theta|\ll 1$ (contrary to that in the QND modules) for realizing an entangled pairs of high fidelity, the single photon detection will be the dominant contribution. 
The maximum success probability of an entangling operation in this situation will be 
\begin{eqnarray}  
P=\frac{1}{2}\left(1-|\langle 0|i\sqrt{2 \eta}~\alpha \sin\theta\rangle\langle \sqrt{2\eta}~\alpha|\sqrt{2\eta}
~\alpha \cos\theta\rangle|\right)\sim \frac{1}{2}\left(1-(2F-1)^{\frac{\eta}{1-\eta}}\right),
\label{E}
\end{eqnarray}
which is half of the optimum success probability \cite{I,D,P} of unambiguously discriminating the coherent state product $|0\rangle_1|\sqrt{2\eta}\alpha\rangle_2$ from $|\pm i\sqrt{2 \eta}~\alpha \sin\theta\rangle_1|\sqrt{2 \eta}~\alpha\cos\theta\rangle_2$. 
As the overlap of these two coherent state products is large in the high fidelity range, there is no contribution from $|0\rangle_1|\sqrt{2\eta}\alpha\rangle_2$, so the success probability in Eq. (\ref{E}) carries a pre-factor $1/2$. Under the present condition, 
the problems to take care in practical implementation are the effieicncy and the dark count of the detector as in any single photon detection.

\section{Conclusion} \label{section5}
In summary, we illustrate a setup that transforms a photon pair in arbitrary rank-four mixed state to a Bell state. 
We realize such transformation with linear optics and Kerr nonlinearity that induces a small XPM phase shift at the single photon level. Using photon number resolving detection, we can realize a near deterministic entangling operation by the setup. Only provided with the simple threshold photon detectors (APDs), its pair generation rate can be as large as the half of the single photon source repetition rate.  
Because the entangled photons are only correlated in polarization modes, such source also enjoys much more flexibility than the other types of entangled pair sources. We also study the optimum performance of the setup in the decoherence environment of photon absorption losses. This universal entangler will be realizable with the current technology if we can stabilize a small XPM phase shift in Kerr media.

\begin{acknowledgments}
B. H. thanks Y.-F. Chen and I. A. Yu for the helpful discussions on the related experimental issues.
This work is supported in part by the Petroleum Research Fund and PSC-CUNY award. 
\end{acknowledgments}

\bibliographystyle{unsrt}

\end{document}